\documentclass[12pt,headings=big,numbers=noenddot,DIV=14,a4paper]{scrartcl}%

\pdfoutput=1

\usepackage{amsmath,amssymb,amsfonts}
\usepackage[normal,font=small,labelfont=bf,labelsep=period]{caption}
\usepackage[pdftex]{color,graphicx}
\usepackage{subfigure}
\usepackage[english]{babel}
\usepackage[compress]{cite}
\addtokomafont{disposition}{\rmfamily\boldmath}

\usepackage[dvipsnames]{xcolor}
\definecolor{darkblue}{rgb}{0,0.2,0.6}
\definecolor{darkgreen}{rgb}{0,0.4,0}

\usepackage[linktoc=page,bookmarks=false,colorlinks=false,
linkbordercolor=RoyalBlue,citebordercolor=ForestGreen,urlbordercolor=CornflowerBlue]{hyperref}

\numberwithin{equation}{section}
\DeclareFontFamily{OT1}{pzc}{}
\DeclareFontShape{OT1}{pzc}{m}{it}{<-> s * [1.10] pzcmi7t}{}
\DeclareMathAlphabet{\mathpzc}{OT1}{pzc}{m}{it}

\title{\vskip 30pt
       \textcolor{black}{Electroweak Corrections}}
\author{\normalsize Riccardo
Barbieri}
\date{\normalsize{\it  Scuola Normale Superiore and INFN, Piazza dei Cavalieri 7, 56126 Pisa, Italy}}

\begin{document}
\begin{titlepage}

\maketitle
\thispagestyle{empty}

\begin{abstract}
\centerline{\bf Abstract}
\noindent
The test of the electroweak corrections has played a major role in providing evidence for the gauge and the Higgs sectors of the Standard Model. At the same time the consideration of the electroweak corrections has given significant indirect information on the masses of the top and the Higgs boson before their discoveries and important orientation/constraints on the searches for new physics, still highly valuable in the present situation.The progression of these contributions is reviewed.
\end{abstract}

\vspace{200pt}

\begin{center}
A contribution to:\\
The Standard Theory up to the Higgs discovery\\
- 60 years of CERN -\\
L. Maiani and G. Rolandi, eds.
\end{center}

\end{titlepage}

\section{Introduction}

In spite of its apparently limited scope the study of the ElectroWeak Corrections (EWC) has played a crucial role in three different directions: 
\begin{itemize}
\item Provide evidence for the consistency at the quantum level of the gauge and the Higgs sectors of the Standard Model (SM);
\item Give significant information, although indirect, on the existence and the mass of the top quark and the Higgs boson before their direct discovery;
\item Constrain and orient the search for possible new physics beyond the SM.

\end{itemize}

Although not making justice to important earlier works on the subject, in order to recall the progressive role of the EWC it makes sense, as we do in the following, to set the $t=0$ time in 1971, when  't Hooft proved the renormalizability of the electroweak sector of the SM\cite{'tHooft:1971fh,'tHooft:1971rn}. Since then, this means more than forty years of progress until the discovery of the Higgs boson, which offers new opportunities for precision physics.

\section{The pioneering works}

A few papers of Veltman in 1977 played a key role both in establishing the program of computing the EWC to physical observables in the SM and in pointing out their possible sensitivity to physics in the 100 GeV range or above, even by measurements of lower energy processes\cite{Veltman:1977kh,Veltman:1977fy}. 

The relevant physical observables are determined at tree level by three parameters: the two gauge couplings in the electroweak sector, $g$ and $g^\prime$, and the vacuum expectation value of the Higgs boson doublet $v$. To fix these parameters, three basic observables may be chosen, which is equivalent to a renormalization procedure. Since the beginning of the electroweak correction program, two such observables have emerged: the electromagnetic fine structure constant, $\alpha$, as measured by the Josephson effect or the electron $(g-2)$, and the Fermi constant, $G_F$, as determined from the muon lifetime. In the seventies  the third observable was more often coming from the $\nu_\mu$-$e $ or the $\nu_\mu$-hadron cross sections, in particular the ratio $R_\nu$ between the neutral and the charged current processes, with a determination, at that time,  of the weak mixing angle $\sin^2{\theta_W}$ at about $10\div 20 \%$ level. The emergence from $\nu$ and $\bar{\nu}$ processes of a $\rho$ parameter close to unity, as predicted at tree level by the SM,   was recognized by Veltman as evidence for the isodoublet nature of the Higgs boson and later, more generally, as a consequence of an approximate {\it custodial symmetry}\cite{Sikivie:1980hm}. At the same time the finding of $\rho$ close to unity
motivated to establish the EWC program.

The absence of any other precisely measured observable, as the same relatively large error on $\sin^2{\theta_W}$ itself from neutrino scattering, could not allow any significant quantitative comparison with the EWC. Yet, as already mentioned, the sensitivity to heavy non-decoupled particles was identified and  relevant one loop effects calculated both due to a split fermion doublet or to the Higgs boson. A first calculated one loop effect was the 
correction to the $\rho$-parameter, the ratio $\rho= G_{NC}/G_F$ between the neutral-current Fermi constant $G_{NC}$, defined in analogy with $G_F$ from the elastic electron-neutrino scattering amplitude at $q^2=0$, and the charged-current Fermi constant itself. The deviation from unity of $\rho$ was calculated from 
a charged lepton of mass $m_l >> M_{W,Z}$, split from a massless neutrino in the same isodoublet\cite{Veltman:1977kh}
\begin{equation}
\rho -1 \equiv \Delta \rho = \frac{G_F m_l^2}{8\pi^2\sqrt{2}}.
\end{equation}
This is the same correction as  for a heavy top, of mass $m_t=m_l$, and a quasi massless bottom up to the extra colour factor of $3$ that has to be introduced.
The neutrino data mentioned above allowed to set an upper bound on $m_l$ of 800 GeV, or -  knowing now that the top is also heavy - of about 450 GeV on the top mass. 

Somewhat later came the calculation in the SM of the two independent one loop corrections growing like the logarithm of the Higgs mass, $m_H$\cite{Passarino:1978jh,Antonelli:1980au,Veltman:1980fk,Sirlin:1980nh,Longhitano:1980iz,Marciano:1980pb}.
As  recalled later on, these corrections affect in a precisely defined way every observable. For example one has
\begin{equation}
\Delta \rho = -\frac{3\alpha}{8\pi\cos^2{\theta_W}}\log{\frac{m_H}{M_Z}},
\end{equation}
\begin{equation}
\frac{\sqrt{2}G_FM_W^2}{\pi\alpha} (1-\frac{M_W^2}{M_Z^2}) \equiv 1 + \Delta r = 1 + \frac{11\alpha}{24\pi\sin^2{\theta_W}}\log{\frac{m_H}{M_Z}}.
\label{deltaR}
\end{equation}
Veltman called "screening theorem" the observation of the mild logarithmic dependence  on the Higgs mass, taken much larger than $M_Z$,  of all the one loop corrections. The root of this theorem is in the absence in the unitary gauge of any coupling of the W and the Z to the Higgs boson proportional to the quartic Higgs coupling $\lambda$.

\section{Constraining $m_t$ and $m_H$}

Although clear evidence for the EWC\footnote{Here and in the following by EWC we mean corrections not of pure electromagnetic nature (nor {\it a fortiori} of strong nature). For the issues discussed here, this  separation  has an operative meaning at a sufficient level of approximation. Needless
 to say the pure electromagnetic (strong) corrections have a very important role in extracting from the experimental data  the physical observables (sometimes called  {\it pseudo-observables} for this very reason). As such these corrections have to be taken into account to a sufficient level of precision.}
 had to wait for the $e^+ e^-$ data at the Z-pole, new more precise measurements came in the eighties from various sources: the direct measurements of the W and Z masses at hadron colliders, $\nu_\mu$-$e$ and $\nu_\mu$-hadron scattering, Atomic Parity Violation and 
$e^+ e^-$ collisions at various energies below the Z-boson mass. This quickly led the focus on the  constraints that could be inferred from these data on the top quark mass and, at least indirectly, on the Higgs boson mass, which was rapidly recognized as relevant  to the constraint on the top mass itself. At the summer conferences in 1989, with the LEP operation just about to start, it was possible to quote the following allowed ranges for the top mass\cite{Ellis:1989uq,Langacker:1989sm}:
\begin{equation}
40~GeV < m_t < 210~GeV~(90\% C.L.)~ for~ m_H < 1~TeV
\end{equation}
\begin{equation}
m_t = 140^{+43}_{-52}~GeV~ for~ m_H = 100~GeV.
\end{equation}
The fit  leading to these constraints had already incorporated as well the fresh first measurement of the Z mass by SLC, $m_Z = 91.17\pm 0.18$ GeV, close to {\it per mille} precision\cite{Abrams:1989aw}. At the same summer conferences UA1, UA2\cite{Gaillard:1989gg} and CDF\cite{Nodulman:1989cm} were quoting lower bounds on the top mass in the $60\div 70$ GeV range.

With the same motivations in mind several calculations were also made, among which:  i) the one loop heavy-top correction to the $Z\rightarrow b \bar{b}$ coupling\cite{Bernabeu:1987me}
\begin{equation}
\delta V_\mu (Z\rightarrow b \bar{b}) = - 2 x \frac{g}{2 \cos{\theta_W}}Z_\mu \bar{b}_L\gamma_\mu b_L,~~
x = \frac{ G_F m_t^2}{8\pi^2\sqrt{2}} ;
\label{Z->bb}
\end{equation}
ii) the two loop heavy-top corrections to the $\rho$-parameter\cite{vanderBij:1986hy}  and iii) the corrections, again to the $\rho$-parameter, growing linearly with $m_H^2$ and  occurring  at two loops\cite{vanderBij:1983bw}. A comprehensive and efficient calculation of all these effects was made possible by the observation that they come from the so called {\it gaugeless limit}  of the SM, through a loop expansion controlled only by the top Yukawa coupling, $\lambda_t$, and by the quartic Higgs boson coupling\footnote{A numerically  relevant exception comes from two loop effects of order $\alpha_S G_F m_t^2$, but they also have nothing to do with the gauging of the electroweak interactions and, as such, are relatively simple to calculate.}\cite{Barbieri:1992nz}.  As an example the corrections to the previously defined quantities, the only ones that grow like powers of $m_t^2$, take a simple form up to two loop order
\begin{equation}
\Delta \rho  = 3 x (1 + x(22-2\pi^2))
\end{equation}
\begin {equation}
\delta V_\mu (Z\rightarrow b \bar{b}) = - 2 x (1+ x(9-\pi^3/3))
\frac{g}{2 \cos{\theta_W}}Z_\mu \bar{b}_L\gamma_\mu b_L.
\end{equation}
These equations hold for $m_H^2/m_t^2 <<1$, but simple closed forms  can be given for any value of $m_H$ and $m_t$. To get these corrections the W and the Z bosons only enter in loops through their Goldstone boson components with couplings controlled by $\lambda_t$ and $\lambda$. Similarly observables with W and Z bosons as external legs are immediately related, via simple Ward identities, to amplitudes for the corresponding Goldstone bosons.

These and other two loop calculations\cite{Fleischer:1992fq,Fleischer:1993ub}, quickly incorporated in programs like ZFITTER, TOPAZ0 or similar, made possible a comparison at the {\it per mille} level  of the SM with a number of precision observables measured at the Z pole  in the early nineties. Consequently the constraint on the top mass became increasingly precise. In the winter conferences of 1994 an indirect determination of the top mass was quoted as\cite{Koratzinos:1994qr,Barbieri:1994cj}
\begin{equation}
m_t = 177\pm 13^{+18}_{-19}~GeV
\end{equation}
with the second error due to the Higgs mass scanning the $60\div 1000$ GeV range. Alternatively one had
\begin{equation}
m_t = 158\pm 11 ~GeV~(m_H = 65~GeV),~~
m_t = 194\pm 10 ~GeV~(m_H = 1~TeV).
\end{equation}
These numbers can be compared with the value of 
\begin{equation}
m_t = 174\pm 10^{+13}_{-12}~GeV
\end{equation}
in the CDF paper, published a  few months later, that provided the first  evidence for direct $t \bar{t}$ production\cite{Abe:1994xt}. 

A pretty analogous story for the Higgs boson mass had to wait twenty years more to be accomplished. In July  2012  the Higgs boson discovery was announced with a mass $m_H = 126.0 \pm  0.4 \pm 0.4$ GeV (ATLAS\cite{Aad:2012tfa}) and $m_H = 125.3 \pm  0.4 \pm 0.5$ GeV (CMS \cite{Chatrchyan:2012ufa}), against an indirect determination from the electroweak fit of $m_H=  97^{+23}_{-17}$ GeV.

\section{Indirect constraints and orientation on new physics}

The indirect sensitivity of the ElectroWeak Precision Tests (EWPT) to physics at higher energies has not only played a role in constraining $m_t$ and $m_H$, but has  also been seen from the beginning as a possible search for signals of new physics. As a matter of fact, until the discovery of at least the top, the separation between the two issues was to some extent not easy to achieve. There are  three different ways that can be used to try to identify the sensitivity of the EWPT to new physics, with relations to each other: i) the examination of effects in specific models; ii) the identification of effective parameters or of effective observables relevant in specific contexts; iii) the use of Effective Field Theory (EFT). For clarity, to describe them it is better not to follow a strict chronological order.

\subsection{Oblique parameters}

Inspired by the early work on heavy Higgs and top in the SM, as already described, and in preparation for the $e^+ e^-$ colliders working at the $Z$ pole, attention was drawn to the contributions to precision observables from the electroweak vacuum polarizations amplitudes\cite{Consoli:1983yn,Kennedy:1988sn}
\begin{equation}
\Pi_{ij}^{\mu\nu}(q^2) = -i\big[ A_{ij}(0) + q^2 F_{ij}(q^2) \big] \eta^{\mu\nu} +(q^\mu q^\nu\mathrm{\small -terms})
\end{equation}
with $i,j = W, Z, \gamma$ or possibly $i,j = 0,3$ for the $B$ or the $W_3$ bosons respectively. Of particular relevance in this context are two quantities $( s=\sin{\theta_W}, c=\cos{\theta_W})$\cite{Peskin:1990zt}
\begin{equation}
\hat{T} = \frac{1}{m_W^2}( A_{33}(0) - A_{WW}(0));\quad
\hat{S} = \frac{c}{s} F_{30}(0).
\label{ST}
\end{equation}
Whenever an expansion in $q^2$ is meaningful, $\hat{T}$ is the leading parameter describing "custodial symmetry" breaking,  since $\hat{T} = \Delta \rho$, whereas $\hat{S}$ is the leading term that respects it. We shall see in  a moment how $\hat{S}$ and  $\hat{T}$ relate to physical observables. 

Although $\hat{S}$ and  $\hat{T}$ are normally defined as due to new physics only, in the SM they are the only terms receiving the one loop logarithms  in a large $m_H$ expansion that we have already encountered,
\begin{equation}
\hat{T} = - \frac{3 \alpha}{8\pi c^2} \log{\frac{m_H}{M_Z}};\quad
\hat{S} = \frac{\alpha}{24\pi s^2} \log{\frac{m_H}{M_Z}}.
\label{TS_logH}
\end{equation}

After electroweak symmetry breaking there are four non vanishing vacuum polarization amplitudes, $\Pi_{WW}, \Pi_{33}, \Pi_{00}$ and $\Pi_{03}$, which, expanded up to first order in $q^2$, are determined by eight coefficients. Three of them are absorbed in the definition of the parameters $g, g^\prime, v$ and two are related by gauge invariance of unbroken electromagnetism: $\Pi_{\gamma \gamma}(0) = \Pi_{\gamma Z}(0)= 0$. As a consequence, other than $\hat{S}$ and $\hat{T}$, it is useful to introduce a third quantity\cite{Altarelli:1990zd}
\begin{equation}
\hat{U} = F_{WW}(0) - F_{33}(0),
\label{U}
\end{equation}
which, however, while also breaking "custodial symmetry", is subleading with respect to $\hat{T}$ in a $q^2$ expansion and, as such, does not contain any "large" $\log{m_H}$ term. At second order in $q^2$ four more parameters appear\cite{Barbieri:2004qk}. Unlike $\hat{S}, \hat{T}, \hat{U}$, two of these four extra parameters respect the electroweak gauge group, but, as $\hat{S}, \hat{T}, \hat{U}$, they are also calculable in the SM, since they are related to  operators of dimension 6.

Let us postpone for a while the determination of $\hat{S}, \hat{T}, \hat{U}$ from experiments.

\subsection{Effective parameters at the $Z$ pole}

As already mentioned, observables at the $Z$ pole have played and are still playing a major role in the comparison between theory and  experiment. This rapidly led to realize that every new physics respecting quark-lepton and flavour universality could affect the $Z$ pole observables and the W-mass only via three effective parameters $\epsilon_i, i=1,2,3$, defined in terms of $\Delta r$ in Eq. (\ref{deltaR}) and of the effective couplings $g_{V,A}^f$ to the $Z$ boson of a fermion $f$ of non zero charge $Q_f$ and third component of the weak isospin $T_{3L}^f$  as\cite{Altarelli:1990zd,Altarelli:1991fk}
\begin{eqnarray}
g_A^f  &=& T_{3L}^f (1+\frac{\epsilon_1}{2}) \label{gA}\\
\frac{g_V^f}{g_A^f} &=& 1 - 4|Q_f| s^2( 1 +\frac{\epsilon_3- c^2 \epsilon_1}{c^2 - s^2})
\label{gVgA}
\end{eqnarray}
\begin{equation}
\Delta r = \frac{1}{s^2}(-c^2 \epsilon_1 + (c^2 - s^2)\epsilon_2 + 2 s^2\epsilon_3)
\label{Dr}
\end{equation}
where
\begin{equation}
s^2 c^2 = \frac{\pi \alpha(M_Z)}{\sqrt{2} G_F M_Z^2}.
\label{s2c2}
\end{equation}
A few comments on these parameters are useful to make.

Eq.s (\ref{gA}) and (\ref{gVgA}) refer to any light charged fermion except the $b$-quark, where already in the SM there is the  flavour non-universal correction (\ref{Z->bb}) due to top loops. To incorporate it together with any other possible new physics effect of the same kind, thus breaking flavor universality, one has to include another  parameter $\epsilon_b$ in the effective couplings of the $b$ quark\cite{Altarelli:1993sz}:
\begin{eqnarray}
g_A^b  &=& -\frac{1}{2}(1+\frac{\epsilon_1}{2})(1+ \epsilon_b)\label{gAb}\\
\frac{g_V^b}{g_A^b} &=& \frac{1}{1+\epsilon_b}(1 - \frac{4}{3} s^2( 1 +\frac{\epsilon_3- c^2 \epsilon_1}{c^2 - s^2}) +\epsilon_b)
\label{eps_b}
\end{eqnarray}

In the parameters $\epsilon_i$ there is no reference to a $q^2$ expansion nor any restriction to vacuum polarization corrections only. An explicit expression for the vacuum polarization contributions to $\epsilon_{1,2,3}$ is  \cite{Barbieri:1991qp}
\begin{eqnarray}
\epsilon_1 &=& e_1 - e_5 + non~oblique \nonumber\\
\epsilon_2 &=& e_2 - c^2 e_5 - s^2 e_4 + non~oblique\\
\epsilon_3 &=& e_3 - c^2 e_5 +c^2 e_4 + non~oblique \nonumber
\end{eqnarray}
where
\begin{equation}
e_1 = \frac{1}{M_W^2}(A_{33}(0)-A_{WW}(0)), \quad
e_2 = F_{WW}(M_W^2) - F_{33}(M_Z^2),\quad
e_3  = \frac{c}{s} F_{30}(M_Z^2), 
\end{equation}
\begin{equation}
e_4 = F_{\gamma\gamma}(0) - F_{\gamma\gamma}(M_Z^2),\quad
e_5 = M_Z^2 F^\prime_{ZZ}(M_Z^2).
\end{equation}
With new physics in oblique corrections only  and a characteristic scale $M_{NP}$ such that an expansion in $M_{W,Z}^2/M_{NP}^2$ is meaningful, a comparison of these equations with eq.s (\ref{ST}-\ref{U}) shows that $\epsilon_{1,2,3}$ may be approximated by $\hat{T}, \hat{U}, \hat{S}$ respectively\footnote{In the SM as an example, although not of new physics, while $\hat{S}$ and $\hat{T}$ contain the leading $\log{M_Z/m_H}$ terms, the full dependence of the EWC on $M_Z/m_H$ is only included in the $\epsilon_i$\cite{Novikov:1992rj,Orgogozo:2012ct}}.

\begin{figure}[t]
\begin{center}
\includegraphics[width=0.44\textwidth]{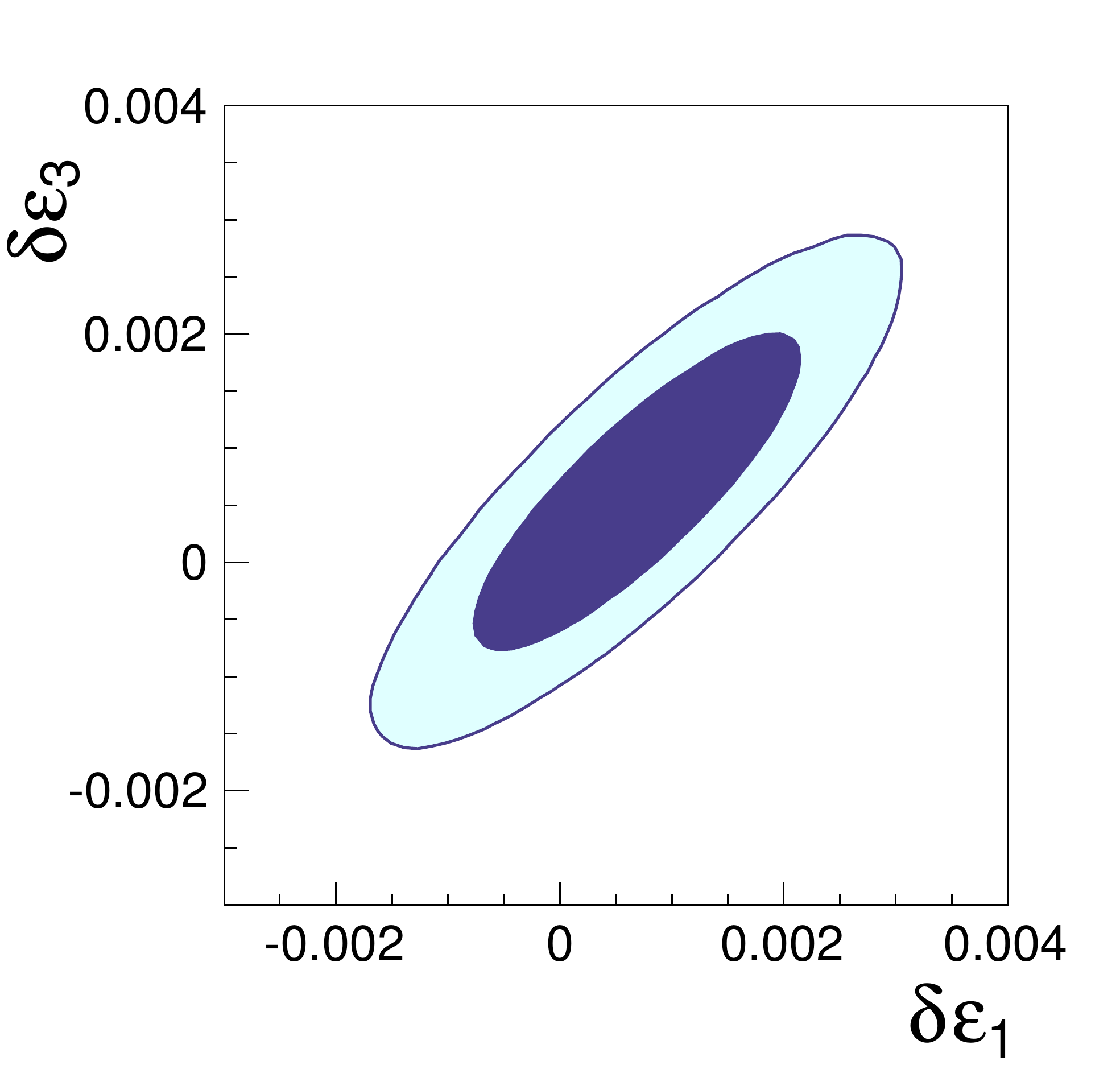}
\includegraphics[width=0.44\textwidth]{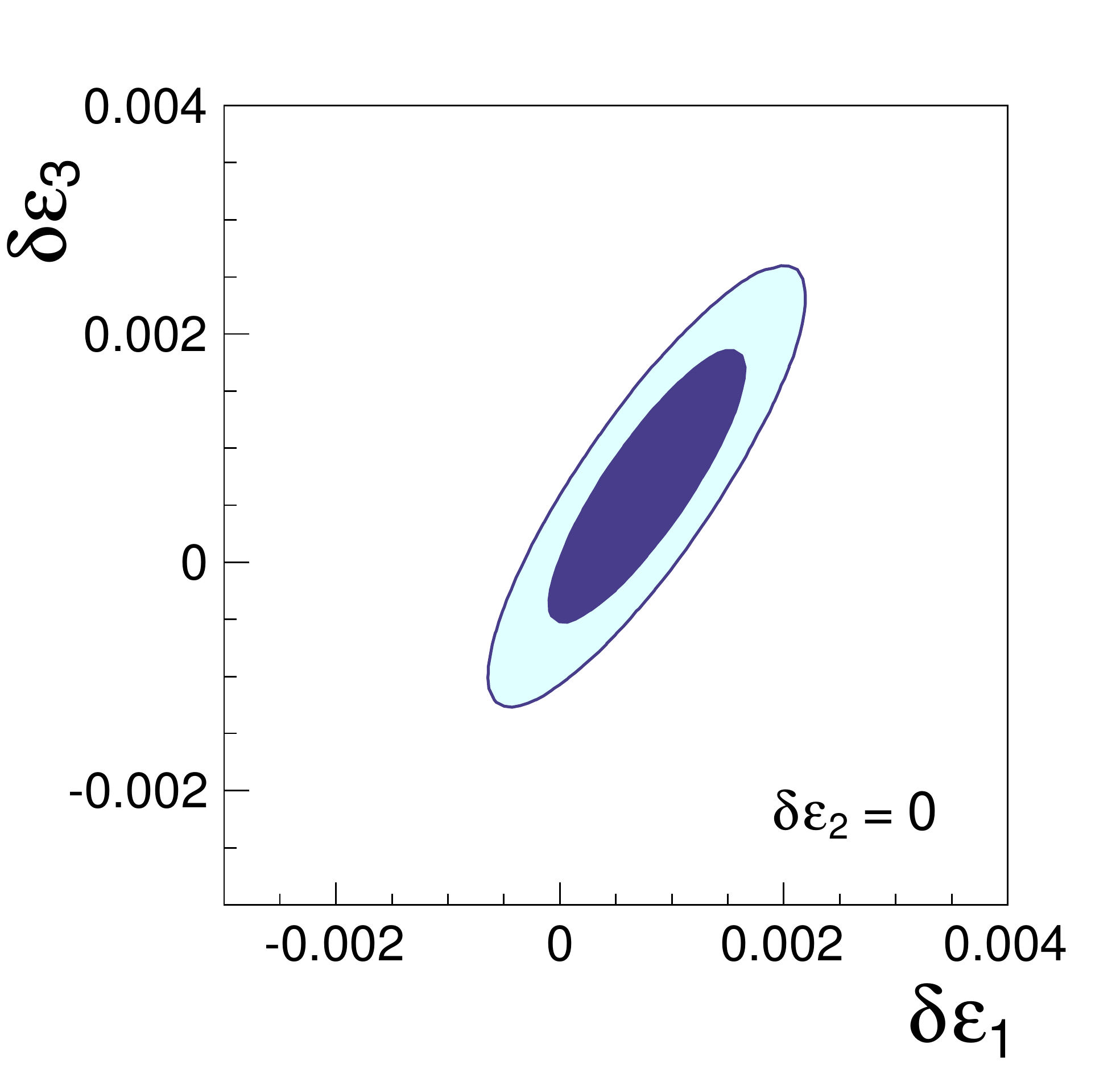}
\includegraphics[width=0.44\textwidth]{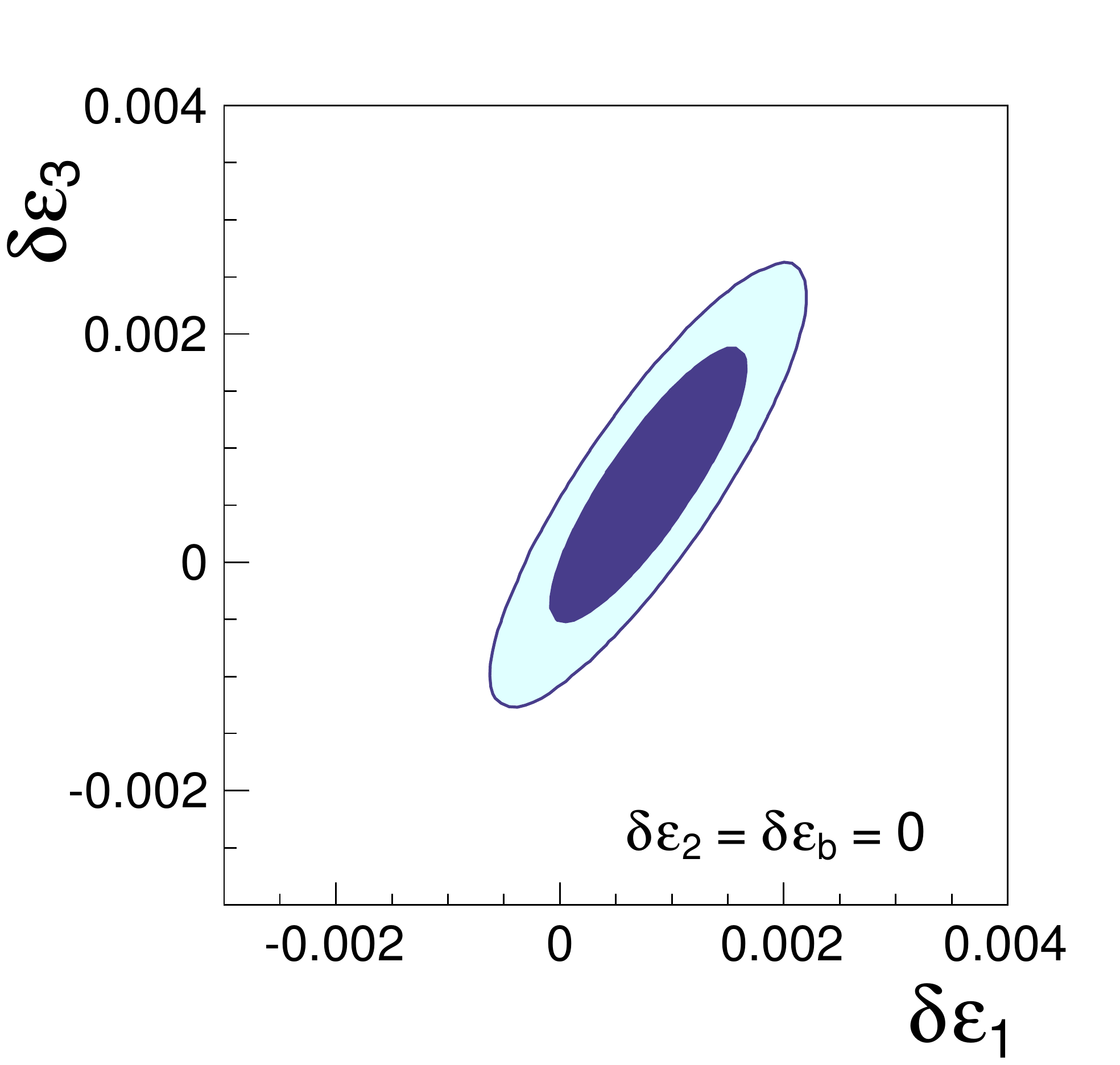}
\caption{\label{running} Two-dimensional probability distributions for $\delta\epsilon_{1}$ and $\delta\epsilon_{3}$ at $68\%$ (the dark region) and $95\%$ (the light region) with all four $\delta \epsilon_{i,b}$ floating (upper left) or fixing $\delta \epsilon_{2}=0$ (upper right)} or $\delta \epsilon_{2}=\delta \epsilon_{b}=0$ (lower). From Ref. \cite{Ciuchini:2014dea}
%
\end{center}
\end{figure}

Note finally that with the definitions (\ref{gA}-\ref{eps_b})
the $\epsilon_i$ include every electroweak correction, both from the SM and from possible new physics, so that it is useful to define the separation $\epsilon_i = \epsilon_i^{SM} (m_t, m_H) + \delta \epsilon_i$.
Now that both $m_t$ and $m_H$ are known with significant accuracy, so are the $\epsilon_i^{SM}$\cite{Ciuchini:2013pca}:
\begin{equation}
\epsilon_1^{SM} = 5.21\cdot 10^{-3},\quad \epsilon_2^{SM} = - 7.37\cdot 10^{-3},\quad \epsilon_3^{SM} = 5.28\cdot  10^{-3}\quad \epsilon_b^{SM} = - 6.9\cdot 10^{-3} .
\label{eps_i_SM}
\end{equation}
This makes possible to plot the two-dimensional probability distributions from current data for $\delta\epsilon_{1,3}$ in Fig.s 1, when all four $\delta \epsilon_{i,b}$ are allowed to float  or fixing some of them to zero\cite{Ciuchini:2014dea}. It depends on a specific model of new physics which one among Fig.s 1 has to be used  to constrain it.

As one can see from these figures, the size of the error on the $\epsilon_i$ is currently at the $10^{-3}$ level, which means, comparing with (\ref{eps_i_SM}), that the electroweak radiative corrections in the SM have been seen at about the $(15\div 20)\%$ level of their typical value. Interestingly this is about the same level of precision at which the loops in Flavour Changing Neutral Current processes are currently tested. So far no deviation from the SM has emerged in either case.

\subsection{Effective operators}

It is customary to view the SM as an effective low energy approximation of a more fundamental theory involving the same low energy degrees of freedom and already valid at shorter distances. As a consequence a more complete Lagrangian describing physics at any scale below a cutoff $\Lambda$ is
\begin{equation}
\mathcal{L}_{eff} (E < \Lambda) = \mathcal{L}_{SM} +  \Sigma_{i, p >0} \frac{c_{i,p}}{\Lambda^p} \mathcal{O}_i^{(4 +p)},
\label{LBSM}
\end{equation}
in terms of any gauge invariant operator $ \mathcal{O}_i^{(4 +p)}$ of dimension $(4+p)$, made of the SM fields.
While the presence of these extra operators has in itself nothing to do with  the EWC, the study of their effects has proven effective in describing the implications of the EWPT. In turn such effects typically arise from  the interference of these operators with radiatively corrected SM amplitudes. This approach has indeed been pursued already in the nineties during the first decade of LEP operation and is receiving further recent attention through the measurements of the Higgs boson properties\footnote{See, among others, Ref.s \cite{Falkowski:2013dza,Pomarol:2013zra,deBlas:2014ula}. }.

The relative drawback of this approach is in the large number of independent operators that are involved and can affect many different observables. Nevertheless it is interesting to recall the most significant bounds on the scale $\Lambda$ that were first obtained after a decade of LEP operation\cite{Barbieri:1999tm,Barbieri:2000gf} and compare them with the corresponding current bounds\cite{deBlas:2014ula}. They are shown in Table 1 at $95\%$ C.L.. The bounds are obtained by considering one operator at a time with the relative coefficient $c_{i,p}=\pm 1$, since what counts is the interference with the radiatively corrected SM amplitudes. Furthermore, in the columns from the data in the year 2000, the Higgs boson mass was taken at 115 GeV, consistent with the direct bound and corresponding to the indirect indication of the EWPT at that time. 
The first operator in Table 1 contributes as well to correct the Higgs boson width into a pair of photons, which is loop suppressed in the SM. From current data on this width one obtains the bounds\cite{deBlas:2014ula}
\begin{equation}
\Lambda > 12.5~TeV\quad (c = -1),\quad\quad
\Lambda > 7.1~TeV\quad (c = 1).
\end{equation}
These are  the strongest bounds obtainable to date from the Higgs boson coupling measurements.

The bounds on $\Lambda$ in the year 2000, compared with the "naturalness" estimate of the radiative corrections to the Higgs boson mass
\begin{equation}
\delta m_H^2 \approx (100~GeV)^2( \frac{\Lambda}{400~GeV})^2,
\label{naturalness}
\end{equation}
and the already mentioned indication for a light Higgs boson in the same year 2000, motivated the title "LEP paradox" of Ref. \cite{Barbieri:2000gf}. Needless to say, the identification of the two $\Lambda$ parameters entering in eq.s (\ref{LBSM}) and (\ref{naturalness}) has to be taken with a grain of salt, since in suitable extensions of the SM
the squares of these parameters are related to each other by a loop factor.

\begin{table}[tb]
\renewcommand{\arraystretch}{1.5}
 \begin{center}
\begin{tabular}{l|cc|cc}
& $c_i=-1$ &  $c_1=+1$ & $c_i=-1$ & $c_i=+1$
\\\hline\hline
$(H^+\tau^aH)W_{\mu\nu}^a B_{\mu\nu}$ & 9.7 & 10 & 11.1 & 18.4
\\
$|H^+D_\mu H|^2$  & 4.6 & 5.6 & 6.3 & 15.4
\\
$i(H^+D_\mu \tau^a H)(\bar{L}\gamma_\mu \tau^a L)$& 8.4 & 8.8& 9.8 & 14.8
\\
$i(H^+D_\mu \tau^a H)(\bar{Q}\gamma_\mu \tau^a Q)$& 6.6 & 6.8 & 9.6 & 8.7
\\
$i(H^+D_\mu H)(\bar{L}\gamma_\mu L)$& 7.3 & 9.2 & 14.8 & 9.2
 \end{tabular}
 \end{center}
\caption{ $95\%$ lower bounds on $\Lambda/$TeV for the individual operators. The first two columns are from the year 2000 data \cite{Barbieri:1999tm,Barbieri:2000gf}, whereas the third and fourth columns are from the currently available data\cite{deBlas:2014ula}.}
\label{tab:u2u3}
\end{table}

\subsection{Examples in specific models}

The Effective Operator approach just described is useful in selecting particular observables not already constrained in an indirect way by other measurements. Nevertheless the large number of independent operators involved and the neglect, inherent to this approach, of any correlation among them may limit the possibility to understand the relevance of the EWC in a particular model of new physics, where, on the other hand, the EWC have been and can be explicitly computed\footnote{In fact caution has to be used in drawing general conclusions about the relative importance of precision in Higgs coupling measurements versus the EWPT, since in specific models of new physics significant correlations may arise between various effective operators that escape a one-by-one operator analysis.  See Ref.  \cite{Barbieri:2013aza} for some relevant examples.}. Let us briefly recall some illustrative examples.

In composite Higgs models or in fact in any model where the standard Higgs boson mixes with a scalar that has anomalous or even vanishing couplings to the intermediate vector bosons, the coupling $g_{HVV}$ of the newly found Higgs boson to the $W$ and the $Z$ will itself deviate from the SM coupling $g_{HVV}^{SM}$. As a consequence, new contributions appear in $\delta \epsilon_{1,3}$ or in $\hat{T}, \hat{S}$ of universal character\cite{Barbieri:2005ri,Barbieri:2007bh}
\begin{equation}
\delta \epsilon_1=- \frac{3\alpha}{8\pi c^2}(1- k_V^2)\log{\frac{\Lambda}{m_H}},\quad\quad
\delta \epsilon_3 = \frac{\alpha}{24\pi s^2}(1- k_V^2)\log{\frac{\Lambda}{m_H}},
\label{eps13kV}
\end{equation}
in terms of
\begin{equation}
k_V =\frac{g_{HVV}}{g_{HVV}^{SM}},
\end{equation}
normally expected to be less than unity, and of a model dependent cutoff $\Lambda$, that roughly measures where $V V$ scattering gets unitarized. The appearance in eq.s (\ref{eps13kV}) of the same coefficients as in eq.s (\ref{TS_logH}) is not accidental. Barring other new physics contributions to  $\delta \epsilon_{1,3}$ of suitable size and sign, which may well be there, a significant constraint is implied on $k_V$ for $\Lambda \gtrsim 1$ TeV by  comparing these equations with  Fig. 1 (in the more appropriate version, depending on the model in question).

The third squark generation, in particular the partners of the left handed top and bottom, are looked for with particular attention since they play a special role in controlling the largest radiative correction to the Higgs mass. Hence the interest in possible radiative correction effects due to virtual exchanges of these particles. An early calculation gave\cite{Barbieri:1983wy,AlvarezGaume:1983gj}
\begin{equation}
\delta \epsilon_1= \frac{G_F m_t^4}{\sqrt{2} ~8\pi^2 m_{\tilde{Q}_3}^2} \approx  10^{-3} (\frac{300 GeV}{m_{\tilde{Q}_3}})^2
\label{T_SUSY}
\end{equation}
where $m_{\tilde{Q}_3}^2$ is the supersymmetry-breaking squared mass for the $(\tilde{t}_L, \tilde{b}_L)$ doublet. Even though eq. (\ref{T_SUSY}) receives corrections from $\tilde{t}_L$-$\tilde{t}_R$ mixing, the sensitivity of $\delta \epsilon_1$ does not exceeds significantly the one in the same eq. (\ref{T_SUSY}) with $m_{\tilde{Q}_3}$ replaced by the lightest stop mass, while it can become weaker. $\epsilon_3$ (or the $\hat{S}$ parameter) receives corrections from the third squark generation which are about one order of magnitude lower than for $\epsilon_1$. Radiative corrections from stop exchanges to $Z\rightarrow b\bar{b}$ and to Higgs couplings may as well play some role in future precision experiments. (See Ref. \cite{Fan:2014axa} for a recent analysis). 

The issue of the corrections to the $g-2$ factor of the muon, with its $3 \sigma$ discrepancy between theory and experiment, is discussed in another contribution to this book\cite{Marciano}.

\section{High precision in the Standard Model}

 \begin{table}[tb]
\renewcommand{\arraystretch}{1.5}
 \begin{center}
\begin{tabular}{l|c|c}
& $\delta M_W/MeV$ & $ \delta \sin^2{\theta_{eff}^l}/10^{-5}$
\\\hline
higher orders &  $5$  & $5$
\\
parametric &  $9$ & $12$
\\
exp. current & $15$  & $16$
\\
exp. FCC-ee & $0.5$  & $0.3$
 \end{tabular}
 \end{center}
\caption{ Theoretical, parametric and experimental errors, present (third line) and future (fourth line), on $M_W$ and $\sin^2{\theta_{eff}^l}$}
\label{tab}
\end{table}

All the content of Sections 3 and 4 rests on the high precision that has been achieved so far in computing the EWC in the SM. Some key quantities like the W mass (or $\Delta r$ in eq. (\ref{deltaR})) and the effective leptonic mixing angle 
\begin{equation}
\sin^2{\theta_{eff}^l} = \frac{1}{4}(1- \frac{g_V^l}{g_A^l})
\end{equation}
have been computed in the SM to full two loop order, which consists of many different contributions\cite{Djouadi:1987gn,Degrassi:1996mg,Degrassi:1996ps,Freitas:2000gg,Awramik:2002wn,Awramik:2002vu,Awramik:2004ge}
 due to the several coupling constants and particles involved, and including as well some relevant  3  loops, going e.g. like $\alpha_S G_F^2 m_t^4$ or $G_F^3 m_t^6$\cite{Faisst:2003px}. As a result the current theoretical uncertainty  from  the missing higher order terms, as inferred for example from comparing the calculations in different renormalization schemes, is estimated  for  $M_W$ at about $ 5$ MeV and on $\sin^2{\theta_{eff}^l}$ at about $5\cdot 10^{-5}$. To these uncertainties one has to add the parametric errors of 9 MeV for  $M_W$ and of $12\cdot 10^{-5}$ for $\sin^2{\theta_{eff}^l}$, obtained with  $\Delta m_t = 1~GeV$, $\Delta \alpha (M_Z) = 3.3\cdot 10^{-4}$ and $\Delta \alpha_S(M_Z)= 7\cdot 10^{-4}$\cite{Degrassi:2014sxa}. This has to be compared with the current experimental errors of
 15 MeV for $M_W$ and $16\cdot 10^{-5}$ for $\sin^2{\theta_{eff}^l}$. All these numbers are collected in Table 2. While the theoretical uncertainties look adequate to compare with current precision measurements, significant improvements will be necessary to make full use of the  precision  foreseen  at future facilities like, e.g., FCC-ee\cite{Gomez-Ceballos:2013zzn}, also shown in Table 2.

\subsubsection*{Acknowledgments}
It is a pleasure to thank G. Altarelli, M. Beccaria, B. Bellazzini, F. Caravaglios, P. Ciafaloni, G. Curci, M. Frigeni, S. Jadach, L. Maiani, A. Pomarol, R. Rattazzi, S. Rychkov, A. Strumia, A. Varagnolo, A. Vicere', for their collaborations on the subject treated in this work. I am indebted with S. Mishima and L. Silvestrini for providing me with Fig,s 1 and to G. Degrassi for help on the uncertainties in Table 2.  This work is supported in part by the European Programme ``Unification in the LHC Era",  contract PITN-GA-2009-237920 (UNI\-LHC) and by MIUR under the contract 2010YJ2NYW-010.
%

\begin{thebibliography}{99}  

\bibitem{'tHooft:1971fh}
  G.~'t Hooft,
  Nucl.\ Phys.\ B {\bf 33} (1971) 173.
  
\bibitem{'tHooft:1971rn}
  G.~'t Hooft,
  Nucl.\ Phys.\ B {\bf 35} (1971) 167.

\bibitem{Veltman:1977kh}
  M.~J.~G.~Veltman,
  Nucl.\ Phys.\ B {\bf 123} (1977) 89.
  
\bibitem{Veltman:1977fy}
  M.~J.~G.~Veltman,
  Phys.\ Lett.\ B {\bf 70} (1977) 253.
  
\bibitem{Sikivie:1980hm}
  P.~Sikivie, L.~Susskind, M.~B.~Voloshin and V.~I.~Zakharov,
  Nucl.\ Phys.\ B {\bf 173} (1980) 189.
  
\bibitem{Passarino:1978jh}
  G.~Passarino and M.~J.~G.~Veltman,
  Nucl.\ Phys.\ B {\bf 160} (1979) 151.
  
\bibitem{Antonelli:1980au}
  F.~Antonelli, M.~Consoli and G.~Corbo,
  Phys.\ Lett.\ B {\bf 91} (1980) 90.
  
\bibitem{Veltman:1980fk}
  M.~J.~G.~Veltman,
  Phys.\ Lett.\ B {\bf 91} (1980) 95.
  
 
  
\bibitem{Sirlin:1980nh}
  A.~Sirlin,
  Phys.\ Rev.\ D {\bf 22} (1980) 971.
  
\bibitem{Longhitano:1980iz}
  A.~C.~Longhitano,
  Phys.\ Rev.\ D {\bf 22} (1980) 1166.
  
  
\bibitem{Marciano:1980pb}
  W.~J.~Marciano and A.~Sirlin,
  Phys.\ Rev.\ D {\bf 22} (1980) 2695
   [Erratum-ibid.\ D {\bf 31} (1985) 213].
%
%
%
%
\bibitem{Ellis:1989uq}
  J.~R.~Ellis and G.~L.~Fogli,
  Phys.\ Lett.\ B {\bf 232} (1989) 139.
  
\bibitem{Langacker:1989sm}
  P.~Langacker,
  Phys.\ Rev.\ Lett.\  {\bf 63} (1989) 1920.
  
\bibitem{Abrams:1989aw}
  G.~S.~Abrams, C.~Adolphsen, R.~Aleksan, J.~P.~Alexander, M.~A.~Allen, W.~B.~Atwood, D.~Averill and J.~Ballam {\it et al.},
  Phys.\ Rev.\ Lett.\  {\bf 63} (1989) 724.
  
\bibitem{Gaillard:1989gg}
  J.~M.~Gaillard,
  Nucl.\ Phys.\ Proc.\ Suppl.\  {\bf 16} (1990) 30.
  
\bibitem{Nodulman:1989cm}
  L.~J.~Nodulman,
  Nucl.\ Phys.\ Proc.\ Suppl.\  {\bf 16} (1990) 40.
  
\bibitem{Bernabeu:1987me}
  J.~Bernabeu, A.~Pich and A.~Santamaria,
  Phys.\ Lett.\ B {\bf 200} (1988) 569.
  
\bibitem{vanderBij:1986hy}
  J.~J.~van der Bij and F.~Hoogeveen,
  Nucl.\ Phys.\ B {\bf 283} (1987) 477.
  
\bibitem{vanderBij:1983bw}
  J.~van der Bij and M.~J.~G.~Veltman,
  Nucl.\ Phys.\ B {\bf 231} (1984) 205.
  

  
\bibitem{Barbieri:1992nz}
  R.~Barbieri, M.~Beccaria, P.~Ciafaloni, G.~Curci and A.~Vicere,
  Phys.\ Lett.\ B {\bf 288} (1992) 95
   [Erratum-ibid.\ B {\bf 312} (1993) 511]
  [hep-ph/9205238];
%
  Nucl.\ Phys.\ B {\bf 409} (1993) 105.
  
  
\bibitem{Fleischer:1992fq}
  J.~Fleischer, O.~V.~Tarasov, F.~Jegerlehner and P.~Raczka,
  Phys.\ Lett.\ B {\bf 293} (1992) 437.
  
\bibitem{Fleischer:1993ub}
  J.~Fleischer, O.~V.~Tarasov and F.~Jegerlehner,
  Phys.\ Lett.\ B {\bf 319} (1993) 249.
  
\bibitem{Koratzinos:1994qr}
  M.~Koratzinos,
  In *La Thuile 1994, Results and perspectives in particle physics* 319-332
  
\bibitem{Barbieri:1994cj}
  R.~Barbieri,
  In *La Thuile 1994, Results and perspectives in particle physics* 375-386
  
\bibitem{Abe:1994xt}
  F.~Abe {\it et al.}  [CDF Collaboration],
  Phys.\ Rev.\ Lett.\  {\bf 73} (1994) 225
  [hep-ex/9405005].
  
\bibitem{Aad:2012tfa}
  G.~Aad {\it et al.}  [ATLAS Collaboration],
  Phys.\ Lett.\ B {\bf 716} (2012) 1
  [arXiv:1207.7214 [hep-ex]].
  
\bibitem{Chatrchyan:2012ufa}
  S.~Chatrchyan {\it et al.}  [CMS Collaboration],
  Phys.\ Lett.\ B {\bf 716} (2012) 30
  [arXiv:1207.7235 [hep-ex]].
  
\bibitem{Consoli:1983yn}
  M.~Consoli, S.~Lo Presti and L.~Maiani,
  Nucl.\ Phys.\ B {\bf 223} (1983) 474.
  
\bibitem{Kennedy:1988sn}
  D.~C.~Kennedy and B.~W.~Lynn,
  Nucl.\ Phys.\ B {\bf 322} (1989) 1.
  
\bibitem{Peskin:1990zt}
  M.~E.~Peskin and T.~Takeuchi,
  Phys.\ Rev.\ Lett.\  {\bf 65} (1990) 964;
  Phys.\ Rev.\ D {\bf 46} (1992) 381.

\bibitem{Barbieri:2004qk}
  R.~Barbieri, A.~Pomarol, R.~Rattazzi and A.~Strumia,
  Nucl.\ Phys.\ B {\bf 703} (2004) 127
  [hep-ph/0405040].
  
\bibitem{Altarelli:1990zd}
  G.~Altarelli and R.~Barbieri,
  Phys.\ Lett.\ B {\bf 253} (1991) 161.
  
 
  
\bibitem{Altarelli:1991fk}
  G.~Altarelli, R.~Barbieri and S.~Jadach,
  Nucl.\ Phys.\ B {\bf 369} (1992) 3
   [Erratum-ibid.\ B {\bf 376} (1992) 444].
 
  
\bibitem{Altarelli:1993sz}
  G.~Altarelli, R.~Barbieri and F.~Caravaglios,
  Nucl.\ Phys.\ B {\bf 405} (1993) 3.
  
\bibitem{Barbieri:1991qp}
  R.~Barbieri, M.~Frigeni and F.~Caravaglios,
  Phys.\ Lett.\ B {\bf 279} (1992) 169.
  
\bibitem{Novikov:1992rj}
  V.~A.~Novikov, L.~B.~Okun and M.~I.~Vysotsky,
  Nucl.\ Phys.\ B {\bf 397} (1993) 35.
  
\bibitem{Orgogozo:2012ct}
  A.~Orgogozo and S.~Rychkov,
  JHEP {\bf 1306} (2013) 014
  [arXiv:1211.5543 [hep-ph]].
  
\bibitem{Ciuchini:2013pca}
  M.~Ciuchini, E.~Franco, S.~Mishima and L.~Silvestrini,
  JHEP {\bf 1308} (2013) 106
  [arXiv:1306.4644 [hep-ph]].
  
\bibitem{Ciuchini:2014dea}
  M.~Ciuchini, E.~Franco, S.~Mishima, M.~Pierini, L.~Reina and L.~Silvestrini,
  arXiv:1410.6940 [hep-ph].
  
\bibitem{Falkowski:2013dza}
  A.~Falkowski, F.~Riva and A.~Urbano,
  JHEP {\bf 1311} (2013) 111
  [arXiv:1303.1812 [hep-ph]].
  
\bibitem{Pomarol:2013zra}
  A.~Pomarol and F.~Riva,
  JHEP {\bf 1401} (2014) 151
  [arXiv:1308.2803 [hep-ph]].
  
\bibitem{deBlas:2014ula}
  J.~de Blas, M.~Ciuchini, E.~Franco, D.~Ghosh, S.~Mishima, M.~Pierini, L.~Reina and L.~Silvestrini,
  arXiv:1410.4204 [hep-ph].
  
\bibitem{Barbieri:1999tm}
  R.~Barbieri and A.~Strumia,
  Phys.\ Lett.\ B {\bf 462} (1999) 144
  [hep-ph/9905281].
  
\bibitem{Barbieri:2000gf}
  R.~Barbieri and A.~Strumia,
  hep-ph/0007265.
  
\bibitem{Barbieri:2013aza}
  R.~Barbieri and A.~Tesi,
  Phys.\ Rev.\ D {\bf 89} (2014) 5,  055019
  [arXiv:1311.7493 [hep-ph]].
  
\bibitem{Barbieri:2005ri}
  R.~Barbieri, T.~Gregoire and L.~J.~Hall,
  hep-ph/0509242.
  
\bibitem{Barbieri:2007bh}
  R.~Barbieri, B.~Bellazzini, V.~S.~Rychkov and A.~Varagnolo,
  Phys.\ Rev.\ D {\bf 76} (2007) 115008
  [arXiv:0706.0432 [hep-ph]].
  
\bibitem{Barbieri:1983wy}
  R.~Barbieri and L.~Maiani,
  Nucl.\ Phys.\ B {\bf 224} (1983) 32.
  
\bibitem{AlvarezGaume:1983gj}
  L.~Alvarez-Gaume, J.~Polchinski and M.~B.~Wise,
  Nucl.\ Phys.\ B {\bf 221} (1983) 495.
  
%
\bibitem{Fan:2014axa}
  J.~Fan, M.~Reece and L.~T.~Wang,
  arXiv:1412.3107 [hep-ph].
  
\bibitem{Marciano}
  W. J. Marciano, this book.
  
\bibitem{Djouadi:1987gn}
  A.~Djouadi and C.~Verzegnassi,
  Phys.\ Lett.\ B {\bf 195} (1987) 265.
  

  
 
  
  
\bibitem{Degrassi:1996mg}
  G.~Degrassi, P.~Gambino and A.~Vicini,
  Phys.\ Lett.\ B {\bf 383} (1996) 219
  [hep-ph/9603374].
  
\bibitem{Degrassi:1996ps}
  G.~Degrassi, P.~Gambino and A.~Sirlin,
  Phys.\ Lett.\ B {\bf 394} (1997) 188
  [hep-ph/9611363].
  
\bibitem{Freitas:2000gg}
  A.~Freitas, W.~Hollik, W.~Walter and G.~Weiglein,
  Phys.\ Lett.\ B {\bf 495} (2000) 338
   [Erratum-ibid.\ B {\bf 570} (2003) 260]
  [hep-ph/0007091];
%
  Nucl.\ Phys.\ B {\bf 632} (2002) 189
   [Erratum-ibid.\ B {\bf 666} (2003) 305]
  [hep-ph/0202131].
  
\bibitem{Awramik:2002wn}
  M.~Awramik and M.~Czakon,
  Phys.\ Rev.\ Lett.\  {\bf 89} (2002) 241801
  [hep-ph/0208113];
  Phys.\ Lett.\ B {\bf 568} (2003) 48
  [hep-ph/0305248].
  
\bibitem{Awramik:2002vu}
  M.~Awramik, M.~Czakon, A.~Onishchenko and O.~Veretin,
  Phys.\ Rev.\ D {\bf 68} (2003) 053004
  [hep-ph/0209084].
  
\bibitem{Awramik:2004ge}
  M.~Awramik, M.~Czakon, A.~Freitas and G.~Weiglein,
  Phys.\ Rev.\ Lett.\  {\bf 93} (2004) 201805
  [hep-ph/0407317].
  
\bibitem{Faisst:2003px}
  M.~Faisst, J.~H.~Kuhn, T.~Seidensticker and O.~Veretin,
  Nucl.\ Phys.\ B {\bf 665} (2003) 649
  [hep-ph/0302275].

\bibitem{Degrassi:2014sxa}
  G.~Degrassi, P.~Gambino and P.~P.~Giardino,
  arXiv:1411.7040 [hep-ph].
  
\bibitem{Gomez-Ceballos:2013zzn}
  M.~Bicer {\it et al.}  [TLEP Design Study Working Group Collaboration],
  JHEP {\bf 1401} (2014) 164
  [arXiv:1308.6176 [hep-ex]].
  
\end{thebibliography}

\end{document}